# Spatial effects on the speed and reliability of protein-DNA search

**Authors**
Zeba Wunderlich[1], Leonid A. Mirny[2,3]

1. Biophysics Program, Harvard University, Cambridge, MA, 02138, USA
2. Harvard–MIT Division of Health Sciences and Technology, Massachusetts Institute of Technology, Cambridge, MA 02139, USA
3. Department of Physics, Massachusetts Institute of Technology, Cambridge, MA 02139, USA

**Abbreviations**
Base pair (bp), one-dimensional (1D), three-dimensional (3D), transcription factor (TF), two-dimensional (2D)

**Abstract**
Strong experimental and theoretical evidence shows that transcription factors (TFs) and other specific DNA-binding proteins find their sites using a two-mode search: alternating between three-dimensional (3D) diffusion through the cell and one-dimensional (1D) sliding along the DNA. We show that, due to the 1D component of the search process, the search time of a TF can depend on the initial position of the TF. We formalize this effect by discriminating between two types of searches: global and local. Using analytical calculations and simulations, we estimate how close a TF and binding site need to be to make a local search likely. We then use our model to interpret the wide range of experimental measurements of this parameter. We also show that local and global searches differ significantly in average search time and the variability of search time. These results lead to a number of biological implications, including suggestions of how prokaryotes achieve rapid gene regulation and the relationship between the search mechanism and noise in gene expression. Lastly, we propose a number of experiments to verify the existence and quantify the extent of spatial effects on the TF search process in prokaryotes.

**Introduction**
Protein-DNA interactions are vitally important for every cell. Transcription factors (TFs) are proteins that interact with specific DNA sequences to regulate gene expression. The targeting of TFs to their sites is a passive process; therefore, it seems natural to assume that TFs simply diffuse through the nucleus (in eukaryotes) or cell (in prokaryotes) until they find their sites.

In the 1970s, this assumption was challenged by the observation that, *in vitro*, the prokaryotic TF LacI is able to find its binding site 100 times faster than expected by three-dimensional (3D) diffusion in the solvent (1). This led to the suggestion of a "facilitated diffusion" mechanism in which TFs alternate between 3D diffusion, *jumping*, through the volume of the cell and one-dimensional (1D) *sliding* along the DNA to rapidly locate their binding sites (2-4). This hypothesis was corroborated by several pieces of evidence – most strikingly several single molecule studies in which the authors visualized individual proteins sliding along DNA (5-7). Several groups have also mathematically modeled this process and shown it to be a plausible way of making the search significantly faster than 3D diffusion (3,4,8-11).

Several aspects of facilitated diffusion, however, remain puzzling, e.g. the effect of the DNA sequence composition and conformational transitions in the protein on the rate of sliding (10,12) and role of the DNA conformation (11). Here we consider how spatial effects influence the search process. Specifically, we ask whether and how search time depends on the initial distance between the protein and the target site.

The distance dependence of the TF search process has not been considered before because the rate of a bimolecular reaction in 3D is distance-independent (13). Therefore, the time it takes for a protein diffusing in 3D to find its target does not depend on the initial distance between the two, as long as this distance is greater than the size of the target. In contrast, the time of search in two dimensions (2D) (e.g. on a membrane) or in 1D (e.g. along DNA or along a filament) is distance dependent (13). Therefore, we ask: can the 1D component of facilitated diffusion make search much faster for proteins starting a small distance from the target site?

Here we use simulations and analytical estimates to demonstrate that TF search time indeed depends on the initial position of the TF with respect to its binding site. We show that the trajectories can be naturally separated into fast *local* and slow *global* searches (Figure 1A). We find that if a TF starts sufficiently close – <1000 base pairs (bp) for our model organism *E. coli* – to its binding site, a local search is likely.

While studying how spatial effects contribute to the search process, we observe that upon dissociation from the DNA, a protein is likely to quickly re-associate near is dissociation point, thus making a short-range *hop,* rather than a long-range *jump* (Figure 1B). We examine how these two types of spatial excursions influence the search process, allowing us to reconcile the widely ranging experimental measurements of the sliding length (6,7,14,15).

Finally, we show that strong non-specific binding of TFs to DNA makes global search rather slow, thus making local search appreciably faster. Moreover, local searches have significantly smaller variance in the search time, making them an attractive mechanism to deliver DNA-binding proteins to their targets quickly and reliably.

There are a number of biological implications of these spatial effects. Since transcription and translation are coupled in bacteria, proteins are produced near the location of their genes. Therefore, TFs whose genes are co-localized with their binding sites are likely to use a local search mechanism. The efficiency of local search provides a physical justification for the observed co-localization of TF genes and their binding sites in prokaryotic genomes (16-18). We also propose a number of experiments to test the mechanism and its predictions.

**Materials and Methods**
*Characterizing hops using simulations*
To include hops in the search model, we needed to estimate the relative frequency of hops and jumps and the displacement due to hops. Assuming that the DNA could be treated as straight rods on the length scale of a hop, we considered the problem in a cylindrical geometry and simplified it further to a 2D geometry (Figure 2A). In the 2D cross-section, DNA strands are represented as absorbing circles. To simulate diffusion in 2D, we discretized the cross section into a 1 μm² square lattice with 1 nm spacing and randomly distributed DNA strands, each with an absorbing radius of 2 nm. We simulated a TF trajectory as a random walk on the lattice, starting from its dissociation from DNA and ending with its association to DNA. Trajectories that started and ended on the same DNA strand were called hops; otherwise they were jumps (Figure 2A). From these trajectories, we calculated the probability of a hop as a function of the number DNA strands in the lattice (Figure 2B).

Using the length of the hop trajectories, we also calculated the displacement along the DNA strand during a hop for lattices with 1500 strands, the approximate density of DNA in *E. coli*. We assumed that, in the 3D geometry, two-thirds of the random walk steps were in the 2D plane and one-third were in the $z$-direction – along the DNA. Therefore, given the length of the hop trajectory in 2D, we drew the number of 1 nm steps along the DNA strand, $z$, from the negative binomial probability distribution function $p(z) = \binom{y+z-1}{z}(2/3)^y (1/3)^z$, where $y$ is the number of 1 nm steps in the 2D cross section. To calculate the net displacement along the DNA strand resulting from a 1D random walk with $z$ 1 nm steps, we drew the displacement (in nm) $c$ from the probability distribution function $p(c) = \frac{1}{\sqrt{2\pi z}} \exp\left[-\frac{c^2}{2z}\right]$, the normal distribution with a mean of 0 and a variance of $z$. As can be seen in Figure 2C, the median absolute displacement resulting from these hops is approximately 1 bp, which is much smaller than persistence length of DNA, the length scale on which DNA is approximately straight, 150 bp. This justified the use of a 2D projection of 3D DNA, since, on the length scale of a hop, DNA is approximately straight. For each DNA density, a total of $10^6$ random walks were simulated, 1000 lattices and 1000 walks per lattice.

*Simulating transcription factor searches*
To simulate the search process in its entirety, we first created a DNA strand $M$ bp long and randomly selected one site to be the binding site. The TF started $d$ bp away from the binding site. (See Tables 1 and 2 for parameter definitions and values.) The TF then alternated between 1D slides and 3D moves (hops or jumps). The slides were modeled as 1D random walks in which the TF could take a 1 bp step to the left or right or dissociate with a probability $p_{dissocation} = 2/s^2$, where $s$ is average number of base pairs scanned during a slide. At the end of each slide, the TF hopped with probability $p_{hop} = 0.8325$ (derived from lattice simulations with 1500 DNA strands), otherwise it jumped. Hops were simulated using the empirical distribution shown in Figure 2C, and jumps were simulated by picking a random association point. When the TF landed on the binding site, the search was terminated.

For the simulation-based estimates of $p_{local}$, the probability that the TF finds its binding site using only slides and hops, but no jumps, we simulated 1000 runs for each combination of $d$ and $s$, using values of $s$ corresponding to $K_d^{NS} = 10^{-6}$, $10^{-5}$, $10^{-4}$ and $10^{-3}$ M, and values of $d$ between 0 and 3000 bp. (See Table 2 for details of the relationship of $K_d^{NS}$, the equilibrium dissociation constant of a TF and piece of non-specific DNA, and $s$.) For the simulated estimates of search time presented in Figures 3B and 3C, we used $K_d^{NS} = 10^{-5}$ M and 5000 runs for each $d$. To find the average search time for $n$ TFs, we simulated runs in groups of $n$, took the minimum search time of the group, and averaged this over all groups.

**Results**

*Why and how is the transcription factor search distance dependent?*
As Polya purportedly told the drunkard wandering the streets looking for his home, "You can't miss; just keep walking, and stay out of 3D!" In 3D, diffusion is non-redundant, i.e. the probability of revisiting a particular site is less than one (13). As a consequence of this property, the average time to find a particular site does not depend on initial position. Conversely, in 1D, diffusion is highly redundant and search time strongly depends on initial position.

In the TF search process, the search time becomes independent of initial position as soon as the TF diffuses in 3D. Therefore, in previous models (3,8-11), the calculated mean search time ($t_s$) is independent of initial position. The search time is presented in different forms, but all are approximately equivalent to the average number of rounds of 1D and 3D diffusion multiplied by the average time of each round:

$$t_s = \frac{M}{s}[\tau_{1D} + \tau_{3D}]. \tag{1}$$

(See Supplementary Material for details.) Here $M$ is the genome length in bp, $s$ is the average number of bp scanned in one slide, $\tau_{1D}$ is the average duration of one slide, and $\tau_{3D}$ is the average duration of jump. (See Tables 1 and 2 for variable and parameter definitions.)

However, if a TF can find its site by sliding along the DNA and not jumping (i.e. "staying out the 3D"), in what we call a *local* search, the search time, $t_s^{local}$, will be dependent on its initial position (Figure 1A). Otherwise, assuming that a jump brings the protein to a random location of the DNA, the search will be *global*, i.e. the TF forgets its initial location and must sample the entire DNA molecule to find its site. In this case, the mean search time, $t_s^{global}$, will be given by equation (1).

Therefore, the mean search time for a TF starting at a distance $d$ from its binding site is an average of $t_s^{local}$ and $t_s^{global}$, weighted by the probability that a TF will find its site via a local search, $p_{local}$, or global search, $1 - p_{local}$:

$$t_s[d] = p_{local}[d] \cdot t_s^{local}[d] + (1 - p_{local}[d]) \cdot t_s^{global} \tag{3}$$

Logically, $p_{local}$ should be a monotonically decreasing function of $d$. Therefore, if a TF starts close enough to its binding site, it is likely to find it using a local search.

*How close does a transcription factor need to be to its site to find it with a local search?*
In the Supplementary Material, we derive $p_{local}$:

$$p_{local}[d] = \exp\left[-\frac{2d}{s}\right] \tag{4}$$

This result is quite intuitive; local searches are likely when a TF starts less than ~$s/2$ bp away from its binding site, half the length covered in a single slide. The sliding length, $s$, depends on a TF's equilibrium dissociation constant for non-specific DNA, $K_d^{NS}$, and its 1D diffusion coefficient, $D_{1D}$, as shown in Table 2. In our model organism, *E. coli*, this value ranges from 30 to 900 bp.

The picture changes when we consider the possibility that some jumps may not be completely randomizing. The current model of jumps assumes that after a TF dissociates from the DNA, all sites on DNA are equally likely to be the association site. However, due to DNA packing, it is likely that there is an increased probability of associating near the dissociation point. Since we do not have a clear picture of DNA packing in the cell, we make the assumption that spatial excursions can be of two extreme types: *hops*, small dissociations from the DNA in which the protein re-associates to the same region of DNA at a distance smaller than or equal to its persistence length (150 bp), and *jumps*, excursions in which each site of DNA is equally likely to be the association point (8,10), (Figure 1B). As we show below, this assumption allows us to study spatial effects on the search process, using only information about the DNA density and its persistence length to characterize hops. Since we do not have enough information to completely characterize jumps, we make the simplest assumption, as others have done (8,10), i.e. all landing points are equally probable.

To include hops into the search model, we first need to estimate the probability of hopping, $p_{hop}$, versus jumping, $p_{jump} = 1 - p_{hop}$. Since we assume that hops happen on length scales shorter than the persistence length of DNA, we can consider the DNA as cylinders, where hopping corresponds to a TF returning to the same cylinder it dissociates from, and jumping corresponds to associating to a different cylinder (Figure 1C). Since we are picturing the DNA as cylinders, we can then move to the 2D problem of return to a circle in the presence of other absorbing points, as depicted in Figure 2A, and use both analytical and simulation-based techniques to estimate $p_{hop}$ using the DNA density in *E. coli*.

To estimate $p_{hop}$ analytically, we make a further approximation by assuming the picture corresponds to two concentric absorbing circles. The inner circle, with radius $R_-$, corresponds to the DNA strand from which the TF dissociates and the outer circle, with radius $R_+$, is an effective shell of absorption by all the other DNA strands. The TF is released at some distance $r$ from the center of the circles. The probability of hopping – returning to the inner circle – is then $p_{hop} = \ln(r/R_+)/\ln(R_-/R_+)$ (13). Since this is only an approximation of the true picture, we use this calculation only to set the bounds for $p_{hop}$ by assuming $R_+$ is minimally the distance between DNA strands, ~ 0.1 μm, and maximally the radius of the *E. coli* cell, ~1 μm. We set $r = 4$ nm and $R_- = 2$ nm, which gives us an estimate of $p_{hop}$ between 0.82 and 0.89. We note that the probability of hopping is still quite high if the TF is released a few nanometers away from the original DNA strand, as newly-translated TFs would be in prokaryotes, where transcription and translation are coupled and TFs are therefore produced in the vicinity of their genes.

Using the same 2D formulation shown in Figure 2A, we also estimate $p_{hop}$ by simulation (Materials and Methods). We find that, for a biological density of DNA, the probability of a TF hopping is large (>0.80). In our subsequent simulations, we assume $p_{hop} = 0.83$, a quantity corresponding roughly to the density of DNA in the *E. coli* nucleoid. In reality, the DNA density is not uniform over the volume of the nucleoid and $p_{hop}$ will vary accordingly. However, in Figure 2B, we show that the change in $p_{hop}$ is small for large changes in DNA density. The obtained value of $p_{hop}$ allows us to calculate the number of hops a TF makes before it jumps to a new region of DNA as $n_{hop} = 1/(1 - p_{hop}) = 1/p_{jump} \approx 6-9$. Using simulations we also find, as others have suggested (3,8), that hops are very short, with a median displacement of 1 bp (Figure 2C). Therefore, in the following treatment, we coarse-grain hops into an effective slide. Thus, the effective sliding time is increased by a factor of $n_{hop} = 1/p_{jump}$ and the effective sliding distance becomes $s_e = s/\sqrt{p_{jump}}$. This gives us:

$$p_{local} = \exp\left[-\frac{2d}{s_e}\right] \tag{5}$$

In *E. coli*, $s_e$ ranges from 70-2000 bp. Figure 3A shows $p_{local}$ as a function of $d$ for several values of $s_e$, as expressed in equation (5) and confirmed by simulation. The correspondence between the simulation, which includes hops explicitly, and the analytical estimate validates our proposal to coarse-grains hops into slides. Thus, the addition of hops simply extends the reach of local searches.

*Why do sliding length measurements vary widely?*
A critical parameter in this analysis is sliding length, $s$. Our analysis may help to understand the wide range in the measurements of sliding lengths for different proteins (see Table 3). Some of these differences are certainly due to differences in the proteins and experimental conditions. In particular, since the non-specific binding of a protein to DNA is driven almost entirely by electrostatics, a protein's non-specific affinity depends strongly on ion concentration (19-21). We also propose that, in some experiments, it is likely that hops are included in the sliding measurement. In the first two experiments listed in Table 3, the experimental designs allow for the unambiguous identification of hops and slides, and the measured slide lengths are on the low end of the scale (14,15).

In the second two experiments, the slide lengths were measured by single-molecule imaging of proteins on DNA (6,7). Given our modeling results, we propose that hops are too short to be seen in a single-molecule experiment. (Halford and Marko also predict this resolution problem (9), though these measurements were not yet made at that point.) The median hop displacement is only 1 bp = 0.34 nm, while the resolution of the experiments is 10-50 nm. Authors of the single-molecule studies have taken the independence of the diffusion coefficient on the ionic strength as an evidence for a lack of hops. Clearly, such small hops could not significantly alter the diffusion coefficient. Our results demonstrate that the major contribution of hops is to duration of sliding rather than to its rate. Thus our model and the notion of small hops help to reconcile these different sliding lengths and seemingly contradicting results about the existence of hops.

*Are local searches much faster than global searches?*
This analysis of local and global searches is not biologically relevant unless there is a significant difference between the length of each, $t_s^{local}$ and $t_s^{global}$. We again use analytical and simulation based approaches. In the Supplementary Material, we show

$$t_s^{local} \approx \frac{d \cdot s_e}{4D_{1D}} \tag{6}$$

$$t_s^{global} \approx \frac{M \cdot s_e}{4D_{1D}} \tag{7}$$

Since $M$, the length of the genome, is quite large compared to values of $d$ for which local search is likely (< 1000 bp), global searches are indeed much longer than local searches.

Figure 3B shows the simulated and estimated values of search time, $t_s$, as a function of $d$ for several different values of $n$, the copy number of TFs per cell. Here we use $K_d^{NS} = 10^{-5}$ M, which gives an effective sliding length of ~700 bp. There is a dramatic difference between $t_s$ for small and large $d$. When considering $n = 10$, the estimated copy number of LacI tetramers per *E. coli* cell (22), the search time of 10 TFs for $d < 700$ bp is less than 3.5 minutes, but is about 15 minutes for $d > 2000$ bp, a time comparable to the duplication time in bacteria.

The initial distance between the TF and its target also affects the reliability of the search. In Figure 3C, we show box-and-whisker plots for the search time of a single TF at $d = 50$, 200 and 2000 bp. Not only is the median dramatically smaller (under 1 second for $d = 50$ and 200 bp compared to over 100 minutes for $d = 2000$ bp), but the spread of the distributions of vastly different – the interquartile range is 0.1 seconds for $d = 50$ bp, 90 minutes for $d = 200$ bp, and 170 minutes for $d = 2000$ bp.

*Why are global searches so slow?*
As has been pointed out by several authors, independent of other parameters, the global search time is minimized when $\tau_{1D} = \tau_{3D}$, i.e. when the TF spends equal amounts of time sliding along the DNA and diffusing through the DNA volume (8-10). This balances the acceleration of the search due to fewer rounds of search with the deceleration due to longer rounds of search. Since

$$\frac{\tau_{3D}}{\tau_{1D}} = \frac{K_d^{NS}}{[D]} \tag{8}$$

where $[D]$ is the concentration of non-specific DNA in the cell, the search time is minimized when $K_d^{NS} = [D]$. In *E. coli*, $[D] = 10^{-2}$ M, and the measured values of $K_d^{NS}$ range between $10^{-3}$ and $10^{-6}$ M (23). Therefore, *in vivo*, $K_d^{NS}$ is not optimized to minimize search time and can result in global search times between 15 and 500 minutes for $n = 1$ TF (Figure 4).

Several other studies that examined the facilitated diffusion mechanism estimated that a TF could find its binding site much more quickly than our estimates of $t_s^{global}$. In their seminal work, Berg, Winter and von Hippel study in some depth the rate of the TF search process (3,4,19). In the concluding paper of a three-paper series, they put together measured and estimated parameters for the search and arrive at a search time of ~2 seconds for $n = 10$ (4). In this estimate, however, they used values of $D_{1D}$ and $D_{3D}$ that are about an order of magnitude larger than recently measured and currently accepted values for *in vivo* diffusion (24,25), and $K_d^{NS} = 10^{-3}$ M, a value at the upper limit of the range. Using our values, we get a search time of ~100 seconds for 10 TFs and ~15 minutes for 1 TF. Three other groups use different approaches to arrive at similar search time expressions. Coppey, et al. use realistic parameter values to estimate a rapid search time for a short piece of DNA, but since they are considering *in vitro* experiments with a restriction enzyme, they do not consider the case where the DNA length is genome-sized (8). In their estimates, Halford and Marko assume that $D_{1D}$ and $D_{3D}$ are equal (and an order of magnitude larger than the measured *in vivo* $D_{3D}$) and that $s$ is optimal, resulting in a rapid search time (9). Slutsky and Mirny also assume $D_{1D}$ and $D_{3D}$ are equal and fast and that $\tau_{1D} = \tau_{3D}$, also resulting in a rapid search time (10).

Since slow global searches are in part due to fairly strong non-specific binding, this naturally leads to the question of why strong non-specific binding would exist. We suggest two possibilities. (i) Strong non-specific binding is functionally important. For example, this binding can be important for relief of repression when a repressor's affinity for its specific site is reduced by ligand binding (23,26). In this case, strong non-specific binding will allow the non-specific sites to out-compete the specific site. For a treatment of other equilibrium aspects of gene regulation, see (27,28). (ii) There is a design limitation. If it is generally true that DNA binding domains use the same set of amino acids to bind both specific and non-specific sites (20), albeit in different ways, there may be a limitation on how weak the non-specific binding can be compared to a strong specific binding.

**Discussion**
In this paper, we examine the distance-dependence of TF search time and find that (i) the search time is distance-dependent, with local searches likely at distances less than sliding length of a TF; (ii) hops lengthen the reach of local searches by increasing the effective sliding length by a factor of ~3; and (iii) due to a TF's strong non-specific affinity for DNA and slow diffusion, global search can be slow. Therefore, low copy-number TFs will find their sites markedly faster if they maintain a small initial distance to their binding sites.

*The role of DNA conformation*
In our model, we attempt to describe the TF search process more realistically by including hops. However, we still assume that jumps are completely randomizing. This assumption is a bit simplistic, though probably sensible, given the data at hand. In reality, the compact conformation of DNA can make jumps non-uniform, e.g. making it more likely for a protein to associate to DNA a certain distance away from a dissociation point (but much further than a hop). For example, the proposed solenoid structure of bacterial DNA can make jumps to the next coil more likely than to a remote coil. Such correlated jumps may make the search more redundant, thus (i) making the global search slower and (ii) making the local search spread further that a single effective sliding length $s_e$. Some have addressed this effect (11,29), and though progress is being made (30), experimental data on the *in vivo* conformation of prokaryotic DNA is still scarce, so it is still unclear what role DNA conformation plays in the search process in live cells. We also note that our work neglects the presence of other DNA-binding proteins that may also interfere with the search process (31).

*Biological implications*
The arrival of a TF to its regulatory site is an essential step in the process of gene regulation. While this step may not necessarily be the rate-limiting one, significant delays in the arrival time can make gene regulation sluggish, thus slowing down response to environmental stimuli and causing the organism to be less fit. We note that these arguments apply to both repressors and activators. A slow search by an activator can lead to delayed gene activation, while a slow search by repressor can lead to unrepressed activity of certain genes or leaky repression. To avoid the adverse effects of slow regulation, we propose that prokaryotes may take advantage of fast local search through a mechanism described below.

Since transcription and translation are coupled in bacteria, proteins are produced *in situ* – near their gene's physical location on the chromosome. We suggest that if a TF gene and its binding site are within $s_e$ bp of each other, this co-localization enables a local search and presumably faster gene regulation. This provides a kinetic advantage that is arguably less costly that maintaining a larger copy number of a TF to compensate for slow search. We believe that strong support for our hypothesis can be found in the organization of prokaryotic genomes. A number of groups have observed that prokaryotic TF genes tend to be closer to their binding sites than expected at random (16,17). An explanation offered is the *selfish gene cluster* hypothesis – the proximity is favorable for horizontal gene transfer of an operon together with its regulator (32,33). Our model offers a kinetic explanation, which is a modified version of Droge and Muller-Hill's idea of "local concentration" (22). In another study, we use bioinformatics to show that TFs with a small number of targets in the genome are likely to be co-localized with their target sites, on length scales comparable to our estimates of $s_e$ (18). We also demonstrate that the observed co-localization and gene orientation cannot be explained by selfish gene hypothesis, further supporting our kinetic hypothesis. For highly pleiotropic TFs with a larger number of target sites, co-localization is impossible, and we suggest rapid search is achieved by high copy number. For example, ArcA, is a highly pleiotropic TF with over 50 binding sites in the *E. coli* genome (34) is estimated to have a copy number of 200 copies per cell (35).

In eukaryotes, where transcription and translation happen in different compartments of the cell, co-localization of this type is clearly not possible. However, eukaryotes have highly organized nuclei, and the compartmentalization may lead to a high concentration of a TF in the vicinity of its binding site (22). Additionally, it appears that some TFs are constitutively bound to their binding sites and await an activation signal (e.g. Gal4 (36)).

Our simulations also demonstrate that a local search has smaller variance of the arrival time. Noise in gene expression is shown to be in part determined by initiation or repression of transcription. Variability in the arrival of a TF to a promoter can greatly increase temporal noise and cell-to-cell variance of gene expression (37-40). Thus cells may employ a local search not only to reduce delays in gene regulation, but also to control (though not necessarily reduce) noise in gene expression.

To estimate the effects of search time on noise, we note that Cai, et. al. have shown that, under the control of a repressor like LacI, protein production occurs in bursts, presumably due to the competition between the repressor and RNA polymerase (41). The frequency of the bursts is proportional to the search time (42). Therefore, the baseline production of a protein that is repressed by a single repressor will scale directly with search time.

*Comparison with a recent in vivo experiment*
A recent *in vivo* single-molecule experiment shows that the 1D/3D search strategy is likely at work in living cells. The experiments studied the search by Lac repressor for its cognate sites. Lac repressor was in its native orientation, i.e. co-localized with the target site, and thus produced at a distance of about 300 bp from the site. The measured search times for one protein per cell were approximately 6 minutes, which is somewhat faster that our estimated global search time if we were to assume $K_d^{NS}$ is $10^{-3}$ M, a value at the very upper limit of the measured *in vivo* range (23). Since the protein synthesis was co-localized with its site, and the YFP marker used had short maturation time of 7 minutes, it is hard to delineate contributions of the local and global search. A more direct test would be to measure and compare the search time for a system where the TF gene is distant from its target site.

*Testing the proposed model with experiments*
We propose a number of ways to test the distance-dependence of the search time. In each case, we propose to compare two strains of *E. coli*, one in which the gene of the TF of interest is less than $s_e$ bp away from its binding site, e.g. within a few hundred bp, and one in which the gene is much farther away, e.g. over 10 kbp away. In the first strain, TFs will be synthesized near their binding sites, making local search likely, and in the second strain, the lack of co-localization will make local search unlikely. Since all the necessary parameters are not known with great accuracy, it is hard to predict the exact differences in search times and the downstream effects between the two strains; however, given the large estimated differences between local and global search times, we would expect the properties measured in the proposed experiments to be detectably different.

First, *in vivo* single molecule measurements (25) can be used to directly measure the binding time in the two strains. Second, one can measure the consequences of co-localization on gene expression by comparing the degree of repression (43), the noise in gene expression (37,38,44) or the dynamics of individual the bursts of expression (45) in strains where the TF of interest represses a reporter gene. Finally, one can compare the more subtle effects of the timing of repression, which are not directly observable but have an impact on fitness. This can be done using competitive growth experiments. One can compete the two strains, both producing a repressor that controls the production of a deleterious protein, but different in relative locations of the repressor gene and its target gene. If our model is correct, the strain with the locally produced repressor will have less leaky repression and therefore a growth advantage over the other strain.

**Funding**
Z.W. is a recipient of a Howard Hughes Medical Institute Predoctoral Fellowship. L.M. is funded by the NEC Research Fund and by the National Center for Biomedical Computing *i2b2*.

**Acknowledgements**
We thank Nickolay Khazanov, Jason Leith, and John Tsang for helpful discussions and the anonymous reviewers for constructive comments on the manuscript.

*Table 1* Model Variables and Functions

| Variable | Description |
|---|---|
| $t_s[d]$ | Average search time for a TF looking for its binding site |
| $t_s^{global}$ | Average search time for a TF looking for its binding site using a global search |
| $t_s^{local}[d]$ | Average search time for a TF looking for its binding site using a local search |
| $d$ | Initial distance between a TF and its binding site |
| $n$ | Copy number of a TF per cell |
| $p_{local}[d]$ | The probability a TF at distance $d$ finds its site using a local search |

*Table 2* Model Parameters and Estimates

| Parameter | Description | Value | Source/Equation |
|---|---|---|---|
| $M$ | Length of *E. coli* K12 DNA multiplied by average copy number | $10^7$ bp | (46) |
| $p_{hop}$ | Probability of a 3D hop | 0.8325 | Simulation |
| $p_{jump}$ | Probability of a 3D jump | 0.1675 | $1-p_{hop}$ |
| $K_d^{NS}$ | Dissociation constant of TF from a nonspecific piece of DNA | $10^{-3}$-$10^{-6}$ M | (23) |
| $k_{on}^{NS}$ | Association rate of TF to a nonspecific piece of DNA; here $b$ = 0.34 nm, the length of a bp of DNA | $10^6$ M$^{-1}$ s$^{-1}$ | Diffusion-limited rate = $4\pi D_{3D} b$ |
| $k_{off}^{NS}$ | Dissociation rate of TF to a nonspecific piece of DNA | $10^0$-$10^3$ s$^{-1}$ | $K_d^{NS} \cdot k_{on}^{NS}$ |
| $[D]$ | Concentration of non-specific DNA binding sites in a cell; here $v_{cell}$ is the cell volume ~ 1 µm$^3$ | $10^{-2}$ M | $\dfrac{M}{v_{cell}}$ |
| $\tau_{1D}$ | Time a TF spends sliding | $10^0$-$10^{-3}$ s$^{-1}$ | $\dfrac{1}{k_{off}^{NS}}$ |
| $\tau_{3D}$ | Time a TF spends hopping and jumping | $10^{-4}$ s | $\dfrac{K_D^{NS}}{[D]k_{off}^{NS}}$ |
| $D_{1D}$ | 1D diffusion coefficient of TF | $1.85 \cdot 10^5$ bp$^2$/s | Mean from (7) |
| $D_{3D}$ | 3D diffusion coefficient of TF | 3 µm$^2$/s | (25) |
| $s$ | Number of bp of DNA scanned by a TF during one slide | 30-900 bp | $2\sqrt{D_{1D}/k_{off}^{NS}}$ |
| $s_e$ | Number of bp scanned in between jumps | 70-2000 bp | $s/\sqrt{p_{jump}}$ |

*Table 3* Recent sliding-length experiments

| Protein | Function | Sliding Distance | Motion Observed | Reference |
|---|---|---|---|---|
| EcoRV | Restriction enzyme | <174 bp | Hopping, sliding | (14) |
| BbvCI | Restriction enzyme | <50 bp | Hopping, sliding | (15) |
| hOgg1 | Base-excision DNA-repair protein | 440 bp | Sliding | (6) |
| LacI | Transcription factor | 350-8500 bp | Sliding | (7) |

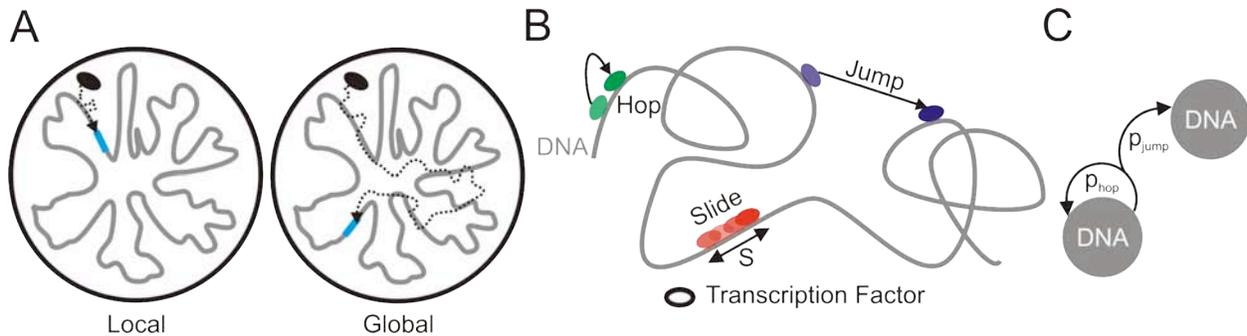

*Figure 1* (A) We defined two types of searches: local searches in which the TF finds its binding site quickly using only hops and slides, and global searches in which the TF finds its binding site using hops, jumps and slides. In this illustration, the black oval is the TF, the gray line is the DNA and the cyan rectangle is the binding site. (*B*) In our model, we consider three types of movements that a TF can make with respect to DNA. *Slides* are rounds of 1D diffusion where the TF remains in constant contact with the DNA for a length of $s$ bp. *Hops* and *jumps* are both types of 3D diffusion. Hops are short, and the dissociation and association sites on the DNA are close (linearly) and correlated. Jumps are long, and the dissociation and association sites may be quite distant along the DNA, though close in 3D space. (*C*) During a search, the TF alternates between 3D and 1D movements until it finds its site. At the end of a slide, the TF dissociates from the DNA, with probability $p_{hop}$ takes a hop and associates to the same strand of DNA, and with probability $p_{jump} = 1 - p_{hop}$ jumps to a new strand of DNA.

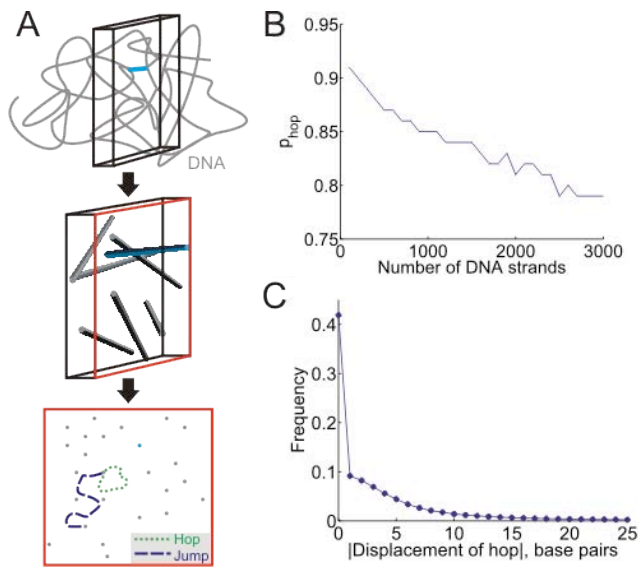

*Figure 2* (*A*) DNA exists in a compacted form *in vivo*, as illustrated on the top. To model the relative frequency and properties of hops and jumps, we looked at a 2D cross section of the DNA, imagining the DNA strands to be approximately straight rods on the short length scales we are dealing with. We defined hops as excursions that begin and end on the same strand of DNA in the cross section, shown with the dotted line, and jumps as excursions that begin and end on different DNA strands, shown with the dashed line. (*B*) Using a lattice model of the cross section, we calculated the probability of hops versus jumps from simulation, using $10^6$ runs. In *E. coli*, the approximate number of DNA strands in the lattice is 1500, which leads to $p_{hop} = 0.83$, but $p_{hop}$ is relatively robust to changes in the DNA density. (*C*) Using the results of the lattice simulation, we calculated the distribution of the displacement along the DNA strand that takes place during each hop.

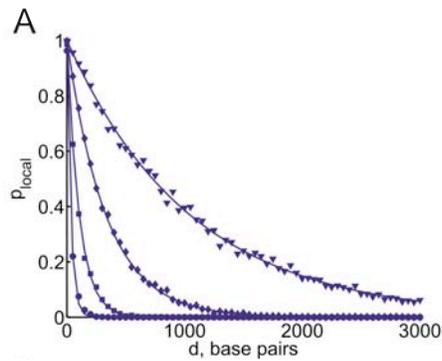

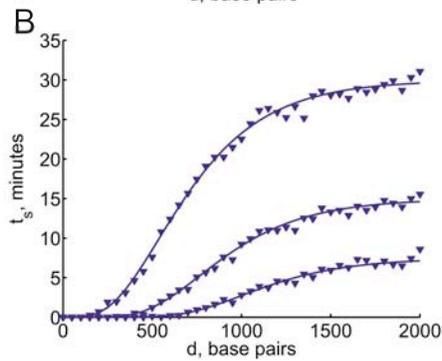

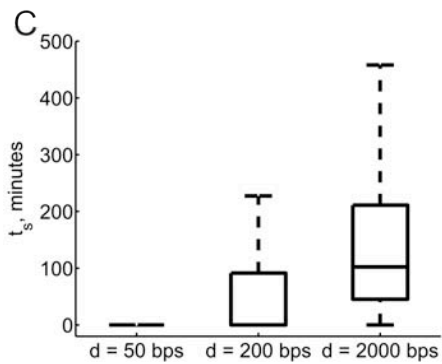

*Figure 3* (*A*) The probability of a local search depends on the effective sliding length, $s_e$, of the TF and the initial distance between the TF and its binding site $d$. Here we show the relationship for several values of $K_d^{NS} = 10^{-6}$, $10^{-5}$, $10^{-4}$ and $10^{-3}$ M corresponding to $s_e = 70$ (circles), 210 (squares), 660 (diamonds), 2100 (triangles) bp, respectively. The solid line represents the analytical result and the markers represent the simulated result ($n_{trials} = $ 1000/condition). (*B*) The average search time $t_s$ depends on several parameters – here we plot it as a function of $d$ for several values of the copy number $n = 5$, 10 and 20 copies/cell; $K_d^{NS} = 10^{-5}$ M. As $n$ increases, the probability of a local search increases and the global search time (the plateau) decreases. For small $n$, the difference in $t_s$ for small and large $d$ is particularly striking. We simulated 5000 runs at each distance $d$. (*C*) The reliability of the search also depends on $d$. Here we plot the distribution of $t_s$ for $d = 50$, 200 and 2000 bp for a single TF. In the box and whisker plots, the box has lines at the lower quartile, median and upper quartile values. The whiskers extend from the box to 1.5 times the interquartile range, the difference between the lower and upper quartiles. Data points beyond the whiskers were excluded.

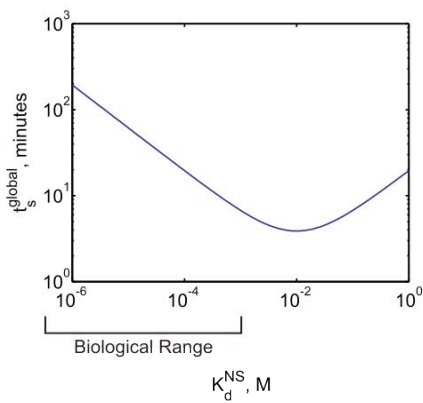

*Figure 4* The global search time for a single TF depends non-monotically on its affinity for non-specific DNA, measured by the dissociation constant, $K_d^{NS}$. The search time is minimized when $K_d^{NS}$ is equal to the concentration of non-specific DNA, $[D] = 10^{-2}$ M. However, the estimated range of $K_d^{NS}$ is $10^{-6}$ to $10^{-3}$ M. See Supplementary Material, Section 1.4.1, for details.